\documentclass[a4paper,11pt]{article}
\pdfoutput=1 

\usepackage{jcappub} 

\usepackage[T1]{fontenc} 
\usepackage{diagbox}
\usepackage{float}
\usepackage{color}
\usepackage{amsfonts,graphics,color}
\usepackage{graphicx}
\usepackage{xcolor}
\definecolor{xlinkcolor}{cmyk}{1,1,0,0}

\newcommand{\redn}{\color{black}}
\newcommand{\black}{\color{black}}
\newcommand{\red}{\black}
\title{\boldmath Towards a model of photon-axion conversion in the host galaxy of GRB~221009A}

\author[a,b]{Sergey Troitsky}

\affiliation[a]{Institute for Nuclear Research of the Russian Academy of Sciences,\\ 60th October Anniversary Prospect 7a, Moscow 117312, Russia}
\affiliation[b]{Faculty of Physics, M.V. Lomonosov Moscow State University,\\ 1-2 Leninskie Gory,  Moscow 119991, Russia}

\emailAdd{st@ms2.inr.ac.ru}

\abstract{GRB~221009A was the brightest gamma-ray burst ever detected on Earth. In its early afterglow phase, photons with exceptional energies \black above 10~TeV \black were observed by LHAASO, and a photon-like air shower \black above $200$~TeV \black was detected by Carpet-2. Gamma rays \black of very \black high energies can hardly reach us from the distant GRB because of pair production on cosmic background radiation. \black Though final results on the highest-energy photons from this GRB have not been published yet, \black a number of particle-physics solutions to this problem were discussed in recent months. One of the most popular ones invokes the mixing of photons with axion-like particles (ALPs). Whether this is a viable scenario, depends crucially on the magnetic fields along the line of sight, which are poorly known. Here, we use the results of recent Hubble Space Telescope observations of the host galaxy of GRB~221009A, combined with magnetic-field measurements and simulations for other galaxies, to construct a toy model of the host-galaxy magnetic field and to estimate the rate of the photon-axion conversion there. Thanks, in particular, to the exceptional edge-on orientation of the host galaxy, strong mixing appears to be natural, both for \black LHAASO and Carpet-2 energy bands\redn, for a wide range of ALP masses $m \lesssim 10^{-5}$~eV and photon couplings $g \gtrsim 10^{-11}$~GeV$^{-1}$.\black}


\arxivnumber{2307.08313}

\begin{document}
\maketitle
\flushbottom


\section{Introduction}
\label{sec:intro}
Opacity of the Universe for high-energy gamma rays, caused by $e^+e^-$ pair production on soft background photons \cite{Nikishov}, is a powerful tool to constrain several exotic particle-physics models or to search for their manifestations, see e.g.\ \cite{ST-mini-rev,Roncadelli-review2022,LIV-review} for reviews. Modification of particle-physics processes may change the standard pattern of pair production, making it possible to observe energetic photons coming from larger distances. One of widely studied examples is the mixing \cite{RaffeltStodolsky} of photons with axions or axion-like particles (ALPs) in the external magnetic field, which may either increase considerably the absorption length of photons \cite{Csaba,Roncadelli-2007}, or allow for conversion of a part of gamma-ray flux into ALPs, which travel unattenuated and reconvert back to photons close to the observer \cite{Serpico,FRT:2009}. For comparison of the two scenarios, see e.g.\ \cite{ST-2scenarios}.

With its unique observational \red appearance\black, the GRB~221009A gamma-ray burst \cite{SWIFT-discovery-paper,GBM-discovery-paper} is suitable for tests of these new-physics hypotheses. Thanks to a relatively nearby, $z\approx 0.151$ \cite{z-GTC-GCN,z-VLT-paper}, location, and to precise pointing of the jet to us, it was the brightest GRB ever observed \cite{BOAT,ordinary,KONUS}. Similar events are expected to be seen by a terrestrial observer once per $\gtrsim 10^4$~years only \cite{BOAT}. Photons of exceptionally high energies associated with the burst have been observed by Fermi LAT up to 99~GeV, and possibly even 400~GeV \cite{FermiLAT-GCN,400-GeV-photon,gamma-Liu,gamma-Laskar,SternTkachev}, and LHAASO WCDA, $\sim 0.3$~TeV to $\sim 5$~TeV \cite{LHAASO-WCDA-GRB}. Strong interest was attracted by \black initial claims of \black events with energies $\approx 18$~TeV by LHAASO KM2A \cite{LHAASO-GCN} and $\approx 251$~TeV  by Carpet-2 \cite{CarpetATel-GRB}. For photons of these energies, one expects suppression of the flux so strong that they could hardly reach the Earth from $z\approx 0.151$ unless some anomalous new-physics effects affected the propagation. Both LHAASO KM2A and Carpet-2 have not yet published details of their observations. 

\black Preliminary KM2A results were mentioned in the talk \cite{LHAASO-ICRC}. The energy of the primary particle for each event is determined in a probabilistic way because of inevitable fluctuations in the shower development. As a result, the estimate of the energy of each particular primary photon depends on the assumed energy spectrum, that is of other photons from the same source. \black The difference between energies reconstructed under different spectral assumptions contributes to systematic uncertainty, because at the highest energies we are interested here, low statistics does not allow to robustly prefer one of the spectral models. \black For the standard log-parabola assumption, the energy of the highest event associated by LHAASO with GRB~221009A was estimated in \cite{LHAASO-ICRC} as $E=17.8^{+7.4}_{-5.1}$~TeV, consistent with the initial 18~TeV estimate. If the assumed spectrum is steeper, the most probable energy, within the uncertainties for a particular event, is lower. \black Other assumptions give $E=12.2^{+3.5}_{-2.4}$~TeV and \black $E=12.5^{+3.2}_{-2.4}$~TeV for the same event \cite{LHAASO-ICRC}. We consider both $E=18$~TeV and $E=13$~TeV in the present work. We will see that this uncertainty does not affect significantly the conclusions of our study\black, changing the range of ALP parameters, for which mixing in the host galaxy is strong, only slightly. Note that this uncertainty is important for interpretation of the observations because of strong energy dependence of the gamma-ray absorption. Implications of these uncertainties for the interpretation may be addressed with full information on the observed spectra, when it is published.\black

Despite uncertainties, \black many theoretical publications on the subject address ALP-related \cite{Roncadelli-newGRB,ST-GRB-JETPL,Meyer-newGRB,Roncadelli-assessment,EW-axion-and-GRB,axion-DM-and-GRB,axion-2210,Marsh,axion-2304,axion-2305}, as well as other \cite{other-2210,other-2211-Smirnov,other-2211,other-2212,other-2301,LIV2210-99GeV,LIV-2210-251TeV,LIV-2210-18TeV,other-2301-Khlopov,LIV-2306-18TeV} new-physics explanations of these anomalies.

\begin{figure}
\centerline{
\includegraphics[width=\linewidth,trim=20 140 20 95,clip]{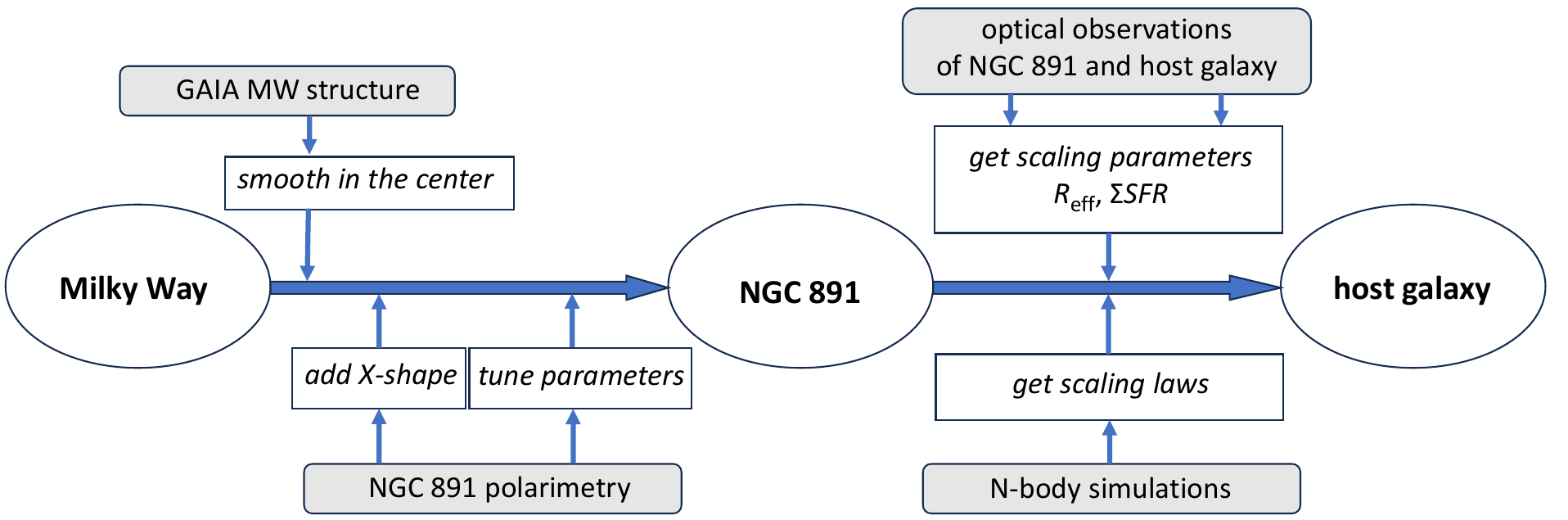}
}
\caption{
\label{f:sketch}
Construction of the toy model of the magnetic field in the host galaxy of GRB~221009A: a schematic view of the procedure. See the text for details. 
}
\end{figure}

Quantitative estimates of the photon-ALP conversion probabilities, necessary \black in particular \black to assess the viability of the corresponding explanation of the anomalous transparency of the Universe for high-energy photons, require knowledge of cosmic magnetic fields along the line of sight. The lack of knowledge of the fields may result in tremendous systematic uncertainties in astrophysical constraints on ALP-photon mixing \cite{LT-magnetic-fields,PDG2022}. For the energies of photons relevant to the GRB~221009A observations by LHAASO KM2A and Carpet-2, significant changes to the standard photon propagation may happen because of magnetic fields in the GRB host galaxy and in the Milky Way. Even for our Galaxy, the magnetic field is known with large uncertainties, see e.g.\ \cite{GMF-Planck-JFnew,GMF-var1,GMF-var2,GMF-var3}, and one should rely on purely theoretical models for the distant host galaxy of GRB~221009A. 

Soon after the observations by \cite{LHAASO-GCN} and \cite{CarpetATel-GRB}, it has been shown that both LHAASO and Carpet-2 events could be explained by ALP-photon mixing, provided this mixing is close to maximal in the host galaxy \cite{ST-GRB-JETPL}. \redn The required ALP parameters are in tension with some model-dependent astrophysical limits, but satisfy constraints from helioscope searches and from stellar evolution, see Sec.~\ref{sec:concl:interp}. \black The assumption of the maximal mixing was reasonably criticized as ``wishful thinking'' by \cite{Roncadelli-assessment}, because nothing was known about the host galaxy for that moment. Assuming that the field in the host galaxy is described by one of the models developed for the Milky Way, and varying the orientation of the host galaxy with respect to the line of sight, \cite{Marsh} claimed that the maximal mixing is not possible for photons of both 18~TeV and 251~TeV simultaneously. 

Newer observations of GRB~221009A by the Hubble Space Telescope (HST) \cite{JWST-HST} brought important information about the host galaxy, and the purpose of the present work is to improve the model of the photon-ALP conversion there by making use of these results. Despite relatively short exposure, Galactic absorption and contamination by the GRB afterglow, these observations severely constrain the conversion model through the galaxy's geometry, the star formation rate, and the position of the GRB in the host galaxy. 
\red
We will show that this information has a drastic impact on the conclusions about the maximal mixing, making them opposite to those of Ref.~\cite{Marsh}, which did not use any observational constraints about the host galaxy.
The maximal mixing is possible for both energies for a certain range of parameters. We however do not address the question whether it can simultaneously explain both LHAASO and Carpet-2 events.

In Sec.~\ref{sec:field}, we construct a toy model of the magnetic field in the host galaxy of GRB~221009A. We start with summarizing relevant observational information in Sec.~\ref{sec:field:obs} and use it to build the model in Sec.~\ref{sec:field:model}, with the help of published observations and simulations of other disk galaxies, including our Milky Way. In Sec.~\ref{sec:ALP-gamma}, we use this model to estimate the photon-ALP conversion rates in the host galaxy, putting the photon source to the GRB. The results are summarized and discussed in Sec.~\ref{sec:concl}. Explicit expressions for the field model of the host galaxy are summarized in Appendix~\ref{a:field} for reference.

\section{Magnetic field of the host galaxy of GRB~221009A} 
\label{sec:field}

\subsection{Observational information about the host galaxy} 
\label{sec:field:obs}
The distance to GRB~221009A was determined by spectroscopic observations of the afterglow \cite{z-VLT-GCN,z-GTC-GCN} and of the host galaxy (HG) \cite{z-host-VLT-GCN,z-VLT-paper}. The redshift is $z=0.151$; assuming the standard cosmology with $\Omega_{\rm M}=0.3$, $\Omega_\Lambda=0.7$ and $h=0.7$, the luminosity distance is $d_L\approx 718$~Mpc and the angular separation of $1''$ corresponds to the projected linear size of 2.63~kpc.

HST observations reveal interesting geometry of HG. Ref.~\cite{JWST-HST} presents parameters of the fit to its image with the S{\'e}rsic profile, effective half-light radius $R_e=(2.45\pm 0.20)$~kpc, S{\'e}rsic index of $1.71\pm 0.18$ and the ratio of half-axes $b/a=0.22 \pm 0.01$. This geometry is remarkable because it corresponds to a rare edge-on orientation of this disk galaxy. We compare it with a nearby (8.4~Mpc) edge-on spiral NGC~891, for which $b/a \approx 0.24$ \cite{NGC891a-b}, and the disk plane is inclined to the line of sight by less than $1^\circ$ \cite{NGC891field-2019}. In what follows, we assume that HG, having even smaller $b/a$, is seen precisely in the disk plane. 

Further, HST images \cite{JWST-HST} allow one to determine the projected position of GRB~221009A in HG. The projected distance from the galaxy center to the GRB is $\approx 0.65$~kpc, with the component parallel to the disk plane of $\approx 0.48$~kpc, and the distance to the middle plane of HG of $\approx 0.44$~kpc. The position of the GRB along the line of sight clearly remains unknown, see Sec.~\ref{sec:ALP-gamma:unknowns}.

According to \cite{JWST-HST}, HG is a typical star-forming galaxy for its redshift, as determined by comparison with other starburst galaxies in general \cite{HST-starforming} and host galaxies of long GRBs in particular \cite{HST-hosts} observed by HST. The HG stellar mass $M_\star$ is modest, estimated as $\log_{10}(M_\star/M_\odot)=9.00^{+0.23}_{-0.47}$ \cite{JWST-HST}.

\subsection{Construction of the toy field model}
\label{sec:field:model}
Further observations with longer exposures, after the GRB~221009A afterglow fades, would bring more information about HG; however, given the distance, there is no hope to obtain direct observational constraints on the magnetic field in the galaxy. First observations, already available, give sufficient information to construct a toy field model which can be used to estimate the probability of the photon-ALP conversion. 

\subsubsection{Procedure}
\label{sec:field:model:general}
The general approach to construct a model of the HG magnetic field, which we follow in detail below, is illustrated by a scheme in Fig.~\ref{f:sketch}. We start with a model of the magnetic field of the Milky Way, which is based on Faraday rotation measurements of numerous radio sources. This model represents well the magnetic field in the spiral arms, but is not reliable for the central part of the Milky Way. We smooth it close to the center, where the Milky-Way spiral structure is not present. Then, we compare it with magnetic-field measurements in NGC~891, a nearby spiral galaxy which has, like HG, higher star formation rate than the Milky Way, and is seen edge-on. We add halo field and tune parameters of the model to reproduce observational data for this galaxy. Finally, we make use of the scaling of the field model with the effective radius $R_e$ of the galaxy and with the surface density of the star formation rate, $\Sigma \mbox{SFR}$, known for both NGC~891 and HG. The scaling relations we use are verified on the set of magnetized disk galaxies obtained in the Auriga simulations.

\subsubsection{Planar field of a spiral galaxy: the Milky Way}
\label{sec:field:model:MW}
Various empirical models of the large-scale Milky-Way magnetic field differ in details but share a general spiral-arm geometry, which is expected to be representative also for fields of other disk galaxies, for a review see e.g.\ \cite{BeckReview}. Our starting point is the model by \cite{Pshirkov-GMF}.  Like other similar models \cite{JF,GMF-Planck-JFnew}, it is not intended to describe correctly the field in the central part of the Galaxy. Indeed, on one hand, the models are heavily based on observations of the Faraday rotation of extragalactic sources, scarce in the directions close to the Galactic Center. On the other, the models were originally developed for studies of deflections of extragalactic charged cosmic-ray particles, arriving isotropically, so that only a small fraction of their flux goes through the central part of the Galaxy. One therefore needs to extend the spiral-arm field model to the central regions of the disk.

Recent analyses of the GAIA data, e.g.\ \cite{BobylevBajkova}, demonstrate that the spiral arms of the Milky Way are traced down to central distances of $\sim 2.5$~kpc. We smooth the field model of \cite{Pshirkov-GMF} at these distances, switching off the spiral-arm structure in such a way that the disk field reaches a value independent of the polar angle in the disk, and then vanishes at the center. Explicit formulae are given in Appendix~\ref{a:field:disk}. Note that this approach is more realistic, compared to simply setting the field to zero in a certain central region, like e.g.\ in the model of \cite{GMF-Planck-JFnew}. However, not being based on observations, this exact functional form of the field in the central region is not guaranteed.

\subsubsection{Matching observations of a nearby edge-on disk galaxy: NGC~891}
\label{sec:field:model:NGC891}
Compared to a typical host galaxy of a GRB, our Milky Way is larger and has lower star-formation rate, cf.\ \cite{more-compact}. In addition, Faraday rotation measurements performed from inside the Galaxy are less sensitive to the magnetic-field component perpendicular to the disk, while this component is important for photon-ALP conversion, given the HG orientation. Indeed, the Faraday rotation is determined by the field component along the line of sight, but the ALP conversion probability is governed by the perpendicular one, see Sec.~\ref{sec:ALP-gamma}. With this motivation, we choose to modify our field model to match observations of a nearby edge-on disk galaxy, NGC~891, already mentioned above.

NGC~891 is well studied thanks to its proximity. Despite being considered as a ``Milky-Way twin'' \cite{Kruit}, NGC~891 is more compact \cite{NGC891a-b} and has the star formation rate about twice higher \cite{NGC891-sfr1,NGC891-sfr2,NGC891-sfr3}, making its properties more similar to those expected for HG.

The magnetic field of NGC~891 was being studied for a long time by different methods \cite{NGC891field-1991,NGC891field-1995,NGC891field-2008,NGC891field-2018,NGC891field-2019,NGC891field-2020}. It was found that, in addition to the plane-parallel field, X-shaped halo field is prominent. Note that stacked observations of halos of various galaxies suggest that the presence of this field component is general \cite{Xshape-general}. At the same time, the halo field in the model of \cite{Pshirkov-GMF} is purely toroidal, see Appendix~\ref{a:field:toro}. We introduce the additional X-shape component to the field model. We use Model~C of \cite{Xshape}; explicit expressions for the field components are given in Appendix~\ref{a:field:Xshape}.

\begin{figure}
\centerline{
\includegraphics[width=0.8\linewidth]{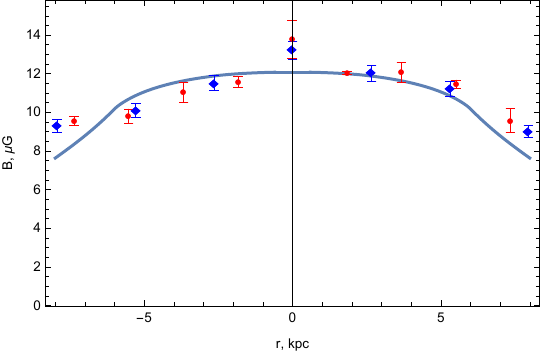}
}
\caption{
\label{f:NGC891-fit}
The profile of the total magnetic field along the central line of NGC~891: our model (line) versus observations by \cite{NGC891field-2018} (red circles) and \cite{NGC891field-2019} (blue diamonds).
}
\end{figure}

The new component is added with an unknown coefficient, and in addition one expects that both the geometrical radial scale and the disk field normalization are different in NGC~891 with respect to the Milky Way. We therefore \red fit \black these three parameters to match observations. To achieve this, we make use of published measurements of the reconstructed magnetic field along the central line of NGC~891 by \cite{NGC891field-2018} and \cite{NGC891field-2019}, which are in a remarkably good agreement with each other, see Fig.~\ref{f:NGC891-fit}.
We average the model field over the polar angle in the disk plane because the orientation and structure of spiral arms is not known. We account for the difference between the regular field $B$, provided by our model, and the total \red equipartition \black field $B_{\rm tot}$, given in Refs.~\cite{NGC891field-2018,NGC891field-2019}, following approximate relations given in Ref.~\cite{BeckReview}, $B \sim \sqrt{3/2} B_\perp \sim 0.3 \sqrt{3/2} B_{\rm tot}$. Note that the turbulent component of the field is marginally relevant for the ALP-photon conversion in the galaxy, see Sec.~\ref{sec:concl:field-var}.  
Because of the polar-angle averaging, the model line is left-right symmetric while the data points are not, reflecting the actual spiral structure. \red The variations in the choice of spiral-arm orientation give additional uncertainty of 14\% (one standard deviation); with this uncertainty added in quadrature to the statistical error bars of individual data points, the goodness of the fit is 96.4\% ($\chi^2 \approx 5.2$ for 12 degrees of freedom). \black
Given rich results of the field measurements, it would be possible to construct a much more precise model for NGC~891, but for our purposes the present toy model is sufficient: we plan to adopt it for HG, for which there are no chance to observe the spiral-arm structure. 

\subsubsection{From NGC~891 to the host galaxy: scaling}
\label{sec:field:model:scaling}
At this final step in the field model construction, we scale the field strength and geometrical parameters to account for the differences between NGC~891 and HG. 

\begin{figure}
\centerline{
\includegraphics[width=0.8\linewidth]{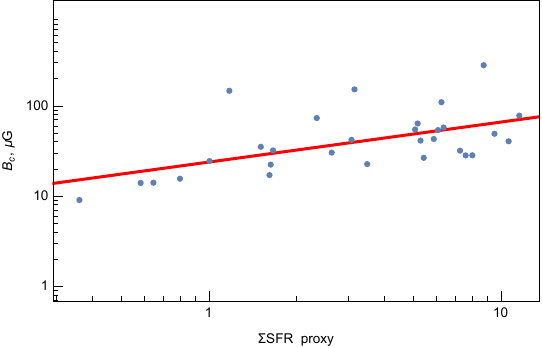}
}
\caption{
\label{f:aufit}
Correlation between specific star formation rate $\Sigma \mbox{SFR}$ and central magnetic field $B_c$ in {\sl Auriga} simulated galaxies \cite{Auriga-galaxies,Auriga-MF}. The proxy for $\Sigma \mbox{SFR}$ is $M_\star/R_e^2$, in units of $10^{10} M_\odot/\mbox{kpc}^2$. The red line represents the best power-law fit $B_c \propto (\Sigma \mbox{SFR})^{0.44}$.
}
\end{figure}

Magnetic field strength increases with the star formation rate (SFR), which can be understood with equipartition arguments, e.g.\ by \cite{equipartition-scaling}, suggesting that the total field scales as $B\propto (\mbox{SFR})^{0.4}$. Other studies established a more pronounced scaling of $B$ with the SFR surface density, $\Sigma \mbox{SFR}$. The proportionality $B\propto (\Sigma\mbox{SFR})^{1/3}$ \cite{scaling-origin-simple} was expected for nearby galaxies; see \cite{scaling-origin-detailed} for a detailed recent discussion and further references. Simple equipartition arguments, see e.g.\ \cite{equip-new}, suggest that $B\propto (\Sigma\mbox{SFR}/v)^{1/2}$, where $v$ is the escape velocity of cosmic rays. The exponent of $1/3$ is reproduced under the assumption of $v\propto (\Sigma\mbox{SFR})^{1/3}$ which is realistic. It is in agreement with observed values of $0.33\pm 0.3$ \cite{scaling-obs2017} or $0.31 \pm 0.06$ \cite{scaling-obs2023}. However, the observations of the magnetic fields are indirect, and the equipartition assumption may be violated in star-forming galaxies \cite{Waxman}, so we test this scaling with the results of detailed cosmological simulations which include the magnetic fields. Figure~\ref{f:aufit}
presents the relation between $M_\star/R^2_e \propto \Sigma \mbox{SFR}$, where $M_\star$ is the total stellar mass and $R_e$ is the effective radius, and the central magnetic field $B_c$ for 30 disk galaxies from the \textit{Auriga} simulations \cite{Auriga-galaxies,Auriga-MF}. They follow the scaling relation $B_c \propto (M_\star/R^2_e)^{0.44 \pm 0.12}$, with the exponent slightly larger than $1/3$. In what follows, we adopt the scaling $B \propto (\Sigma \mbox{SFR})^{0.4}$ and note that the effect of the changes in the exponent on our results is negligible. We use this rule to scale the field strengths of all three components of our model from NGC~891 to HG. For the former, we use the value of $\Sigma \mbox{SFR}\approx 3.13 \times 10^{-3}\, M_\odot/\mbox{kpc}^2/\mbox{yr}$ \cite{NGC891-sfr3}. For HG, we use the fact that it is a typical star-forming galaxy for its redshift \cite{JWST-HST} and invoke the relation between $M_\star$ and $\mbox{SFR}$ from \cite{Popesso}, making use of Eqs.~(15), (12) and Table~2 of that reference. Then we estimate $\Sigma \mbox{SFR} \sim \mbox{SFR}/(2\pi R_e^2)$. Here, $1/2$ comes from the fact that $R_e$ is the half-light radius. For $z=0.151$ and $M_\star=10^9 M_\odot$, we obtain $\Sigma \mbox{SFR}\approx 8.5 \times 10^{-3}\, M_\odot/\mbox{kpc}^2/\mbox{yr}$. 

Additionally, we need to change all linear scales\red, listed in Table~\ref{tab:param}, \black in the model to reflect the difference in the galaxy sizes. It is natural to use $R_e$ as the scale parameter. We tested with the \textit{Auriga} galaxies that the linear scaling is reasonable, though with a large scatter\red: the scale of the magnetic field decrease, $r_{0,\rm inner}^B$, determined from simulations, is proportional to $R_e^{0.88 \pm 0.35}$\black. We use $R_e=5$~kpc for NGC~891, consistent with observations by \cite{Kruit} ($R_e\approx 4.9$~kpc) and by \cite{NGC891-Re-1998} ($3.93~\mbox{kpc} <R_e<5.80$~kpc). For HG, we use $R_e=2.45$~kpc determined by \cite{JWST-HST}.

This rescaling finalizes the construction of the toy model of the HG magnetic field. Explicit expressions for the field components and values of parameters for HG are summarized in Appendix~\ref{a:field} for reference.

\red Figure~\ref{fig:Bplot}
\begin{figure}
\centerline{
\includegraphics[width=0.8\linewidth]{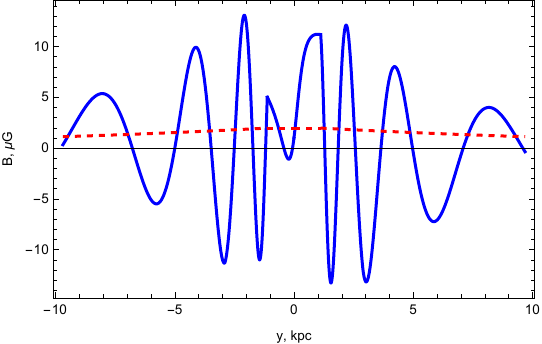}
}
\caption{
\label{fig:Bplot}\red
Model HG magnetic field components perpendicular to the line of sight from the Earth to the GRB, as a function of the coordinate $y$ along this line. $\theta_0=80^\circ$; $y=0$ corresponds to the middle line of the galaxy. The blue full line represents the component in the disk plane, the red dashed line is the component perpendicular to the disk. 
}
\end{figure}
illustrates the magnetic field components relevant to the photon-ALP conversion discussed in the next section.
\black

\section{Photon--ALP conversion probability}
\label{sec:ALP-gamma}
\subsection{Formalism}
\label{sec:ALP-gamma:form}
The mixing of photons with light pseudoscalar particles in the external magnetic field underwent extensive theoretical studies since the original works  \cite{Maiani,RaffeltStodolsky}, for reviews in the astrophysical context see e.g.\ \cite{ST-mini-rev,Roncadelli-review2022}. A general ALP has two parameters, mass $m$ and photon coupling $g$ (we follow the notations of \cite{LT-magnetic-fields}). In particle-physics units $\hbar=c=1$, both $m$ and $1/g$ have the dimension of energy $E$. The easiest and numerically fast way to estimate the probability of conversion of a photon to an ALP is to introduce the $3\times 3$ density matrix $\rho$ and to find its evolution with the line-of-sight coordinate $y$ described by the equation
\begin{equation}  
 i \frac{d\rho(y)}{dy}=\left[\rho(y),\mathcal{M}(E,y)  \right],
\label{Eq:rho-eq}
\end{equation}
where
\begin{equation} 
\mathcal{M}=\frac{1}{2}
\left(
\begin{array}{ccc}
\red \Delta_1 \black & 0 & igB_{1}\\
0 & \red \Delta_2 \black & igB_{2}\\
igB_{1} & igB_{2} & m^{2}/E
\end{array}
\right),
\label{Eq:M}
\end{equation}
$B_{1,2}(y)$ are two components of the magnetic field perpendicular to the line of sight\red, and $\Delta_{1,2}$ are the elements describing photon dispersion. A convenient expression for $\Delta_{1,2}$ reads (see e.g.\ \cite{FRT:2009,Raffelt1,Kachelriess}),
\begin{equation}
\Delta_{1,2}\approx \left(8.0 +1.43\left(B_{1,2}^2 +\frac{4}{7} B_{2,1}^2\right) \right)\times 10^{-8} \, \mbox{pc}^{-1} \, \frac{E}{\mbox{TeV}}.
    \label{Eq:dispersion}
\end{equation}
Here, the first, constant, term in parentheses is due to the interaction with CMB photons \cite{Raffelt1}, while the second one, proportional to $B^2$, comes from the QED vacuum polarization \cite{Adler,Ritus}. The effect of $\Delta_{1,2}$ grows with energy but, even for $E=251$~TeV, remains small provided
\begin{equation}
1.3 \left( \frac{g}{10^{-11}~\mbox{GeV}^{-1}} \right)^{-1}
<
\frac{B}{\mu \mbox{G}} 
<
4.2 \, \frac{g}{10^{-11}~\mbox{GeV}^{-1}} .
\label{Eq:dispersion-small}
\end{equation}
Thanks to the exceptional orientation of both HG and the Milky Way, the line of sight crosses several spiral arms, and $B$ changes considerably along it. As a results, the conditions (\ref{Eq:dispersion-small}) are satisfied, for a range of $g$, at least for a fraction of the line of sight, sufficient for the photon-ALP conversion. In the numerical calculations, $\Delta_{1,2}$ were included.
\black In the expression \red(\ref{Eq:dispersion})\black, we kept only the terms which are not negligible for the parameter ranges we are interested in here. Complete expressions for other terms, \red which include the plasma-related photon dispersion and terms in the vacuum dispersion suppressed by $\kappa=(E/m_e)(B/B_{\rm cr})$, where $m_e$ is the electron mass and $B_{\rm cr}=m_e^2/e\approx 4 \times 10^{13}$~G, \black can be found e.g.\ in \cite{FRT:2009}. \redn The plasma term may be neglected provided the electron density $n_e$ satisfies
\begin{equation}
    n_e \ll 10^5\,\mbox{cm}^{-3}\, \frac{B}{\mu \mbox{G}} \, \frac{g}{10^{-11}~\mbox{GeV}^{-1}} \,\frac{E}{\mbox{TeV}},
\end{equation}
which is safely fulfilled even in the star-forming regions. Smallness of higher-order terms in $\kappa$ is guaranteed by the condition
\begin{equation}
    \kappa \approx 1.8 \times 10^{-14} \, \frac{B}{\mu \mbox{G}}\,\frac{E}{\mbox{TeV}} \ll 1.
\end{equation}
\black

The assumption that the emitted photons are not polarized is equivalent to the initial condition
\begin{equation}
\rho(\black y_0\black)=\mbox{diag}\left(1/2,1/2,0\right).
\label{Eq:rho0}
\end{equation}
Then, the probability to observe an ALP
at a distance $y$ from the source is given by the
$\rho_{33}(y)$ component of the solution $\rho(y)$ to
Eq.~(\ref{Eq:rho-eq}) with the boundary condition (\ref{Eq:rho0}). 

\subsection{Unknown unknowns}
\label{sec:ALP-gamma:unknowns}
For a given energy and fixed ALP parameters, the conversion probability is determined by the evolution of $B_{1,2}$ along the line of sight. We will use the toy model developed in Sec.~\ref{sec:field:model} to estimate this probability for HG. However, even if the model were precise, it would represent an unrealistic task to determine two important parameters. One is the orientation of the spiral arms in the disk plane, mentioned already in Sec.~\ref{sec:field:model:NGC891}, parameterized by the zero point \black $\theta_0$ \black of the polar angle $\theta$ in expressions of Appendix~\ref{a:field}. Another is the deprojected location of the GRB site in HG, which determines the zero point \black $y_0$ \black of the coordinate $y$ along the line of sight, where the condition (\ref{Eq:rho0}) should be applied. 

To tackle this issue, we consider a variety of possible values for the both parameters and perform calculations for different assumptions. \textit{A priori}, all orientations of spiral arms have equal probabilities, while the position of the GRB along the line of sight is expected to follow the stellar distribution. To estimate the latter, we use parameters of the S{\'e}rsic profile of HG reported by \cite{JWST-HST} and reconstruct the deprojected luminosity distribution following \cite{Sersic-deproj1,Sersic-deproj2}. 

The calculations are performed as follows. For each pair of $(m,g)$ on a logarithmic grid, $-9\le\log_{10}(m/\mbox{eV})\le-5$, and $-12\le\log_{10}(g/\mbox{GeV}^{-1})\le-9.5$, with the step of 0.1 in both cases, and for each of the two energies, 18~TeV and 251~TeV, we generate 100 pairs of random numbers $(\theta_0,y_0)$, where the uniformly distributed $0\le\theta_0<2\pi$ is the zero point of $\theta$ with respect to the direction perpendicular to the line of sight, and $y_0$ is the GRB location along the line of sight, distributed following the deprojected luminosity distribution in HG. Fixing the field geometry by $\theta_0$ and the GRB location by $y_0$, we calculate the probability for a photon to convert to ALP in HG for this case. 

\subsection{Results}
\label{sec:ALP-gamma:results}
For each point in the parameter space, we thus obtain a distribution of the probabilities for an energetic photon emitted from GRB~221009A to convert to an ALP in HG, reflecting the freedom in $\theta_0$ and $y_0$. To formulate the results, we select the parameters for which the probability exceeds 32\% (maximal mixing) or 10\% (strong mixing) for at least 68\% of realizations of $(\theta_0,y_0)$. 

\begin{figure}
\centerline{
\includegraphics[width=0.8\linewidth]{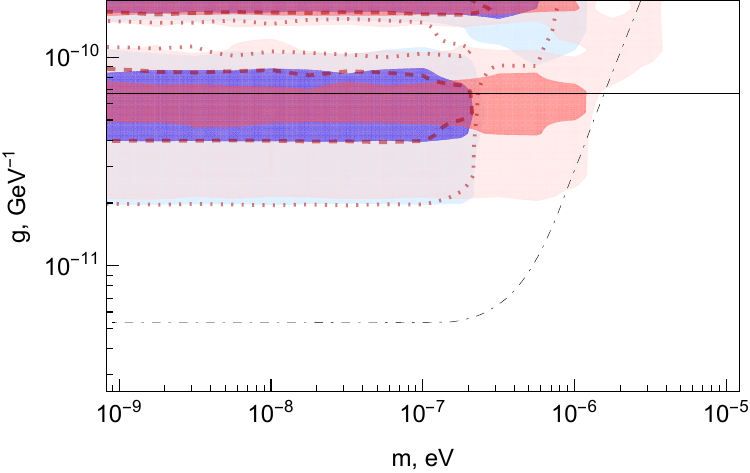}
}
\caption{
\label{f:probplot}
The ALP parameter space: mass $m$ and photon coupling $g$. For parameters in the shaded regions, maximal (conversion probability $>32\%$, dark shading) or strong (conversion probability $>10\%$, light shading) photon-ALP mixing in the host galaxy occurs for $>68\%$ of geometry realisations, see the text. Blue shading corresponds to $E=18$~TeV, pink shadings -- to $E=251$~TeV. In the darkest regions, they overlap and the maximal mixing happens for both energies.
\black Dashed and dotted lines delimit the regions of maximal and strong mixing, respectively, assuming $E=13$~TeV. \black
The full line represents the upper limit on $g$ from the search of solar ALPs in the CAST experiment \cite{CAST}, coinciding with the limit from the evolution of horizontal-branch stars in globular clusters \cite{HBstars}.
The dash-dotted line represents the upper limit on $g$ from magnetic white dwarf polarization \cite{MWD-polarization}, the strongest of astrophysical limits which assume a particular magnetic-field configuration in the source. Purely laboratory experiments give limits too weak to be shown in this plot, see \cite{PDG2022} for their compilation.
}
\end{figure}

The results are presented in Fig.~\ref{f:probplot}.
For a wide range of ALP parameters $m$ and $g$, strong mixing in HG is possible both for $E=18$~TeV and $E=251$~TeV. \black The same is true for somewhat lower energies, like $E=13$~TeV, obtained for the highest-energy LHAASO event under different assumptions. \black  As it is expected from simple estimates, e.g.\ \cite{FRT:2009}, efficient conversion at higher energy, 251~TeV, is possible for higher ALP masses; we note that the conversion probability for 18-TeV photons there is lower but not negligible. 
\red
The complicated structure of the plot reflects the fact that for strong mixing, for fixed distance of the source and for a given magnetic field, the conversion probability is an oscillating function of the ALP parameters. Additional complications arise from the structure of the spiral-arm field and the procedure of sampling the ``unknown unknowns''.
\black

\begin{figure}
\centerline{
\includegraphics[width=0.8\linewidth]{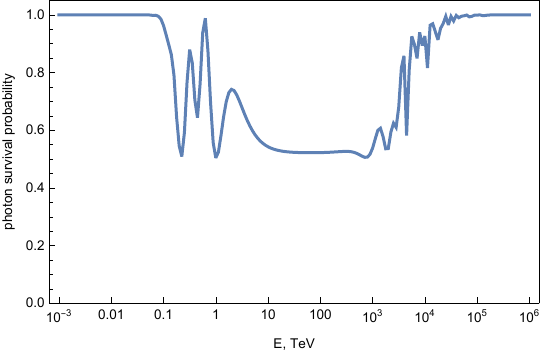}
}
\caption{
\label{fig:suppfac}\red
The energy dependence of the photon survival probability at the exit of HG, assuming $m=10^{-7}$~eV, $g=6\times 10^{-11}$~GeV$^{-1}$, $y_0=0$, $\theta_0=80^\circ$. Absorption in HG is neglected.
}
\end{figure}

\red
Figure~\ref{fig:suppfac}
presents the energy dependence of the photon survival probability at the exit from HG for a particular realization, $\theta_0=0$ and $y=80^\circ$, of the magnetic field. It demonstrates that, indeed, the maximal mixing is possible for the energy range which includes both 13~TeV and 251~TeV. At higher energies, dispersion effects suppress the mixing.
\black

\section{Discussion and conclusions}
\label{sec:concl}
\subsection{Variations in the field model}
\label{sec:concl:field-var}
We have demonstrated that, within the particular toy model of the HG magnetic field we constructed, strong photon-ALP mixing is possible in HG for photon energies corresponding to observations of GRB~221009A by both LHAASO KM2A and Carpet-2. While the field model reflects the best up-to-date knowledge about HG, variations in the field strength and configuration may change the conversion probabilities. These variations may come from several sources.

Firstly, our starting point was a model of the Milky-Way magnetic field, which is also not firmly known. Particular models give different results, especially considering the field outside the disk, but also regarding the disk field normalization. However, these differences are not relevant for our study because we take only the shape of the disk field from this model, and subsequently change both the field strength and geometric scales as described in Sec.~\ref{sec:field:model:NGC891}, \ref{sec:field:model:scaling}. The halo field is also added independently, Sec.~\ref{sec:field:model:NGC891}. Therefore, our field model essentially inherits only the overall spiral-arm structure from the original Milky-Way field, which is based on the Galaxy structure and does not vary between models. This makes the result insensitive to the choice of the original model, so we safely use the model of \cite{Pshirkov-GMF}, for which simple analytical expressions are available, see Appendix~\ref{a:field:disk}.

Secondly, we discuss here only the regular large-scale magnetic fields, while the total field strength in galaxies includes non-negligible turbulent component. It is especially strong in regions of intense star formation \cite{MF-SFregions}, which are plausible sites of GRBs \cite{HST-hosts}. However, the fields with coherence lengths $\lesssim 100$~pc, which is a typical size of a star-forming region, have very little effect on the ALP conversion: the turbulent field is of a similar order as the regular one, or a factor of a few stronger, \red but its relatively rapid change does not allow the conversion to build up coherently\black. This results in a subleading correction to the total conversion probability, as can be seen from e.g.\ \cite{FRT:2009}, and justifies the ignorance of the turbulent galactic field in this and other studies. \red As it is demonstrated e.g.\ in \cite{turbulent}, the turbulent field becomes important for the shape of the spectrum at energies between low and strong mixing. \black For a detailed quantitative calculation of the effects of the ALP conversion on the observed photon spectra, the turbulent fields may be taken into account, provided their parameters are known robustly. However, they could hardly affect the answer to the main question of the present paper, whether or not the strong mixing in HG is possible.
\begin{figure}
\centerline{
\includegraphics[width=0.8\linewidth]{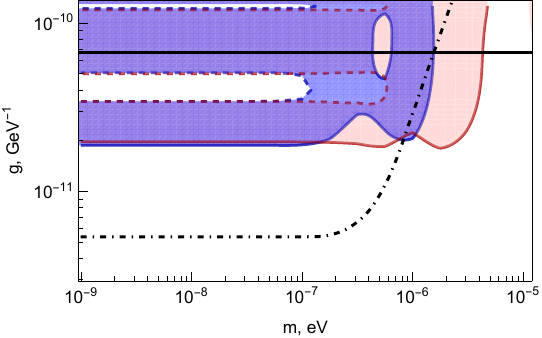}
}
\caption{
\label{f:HG-MW}
\black
ALP parameters (mass $m$ and photon coupling $g$) for which strong mixing happens in the Milky Way (regions limited by dashed lines, direction to GRB~221009A, field model \cite{Pshirkov-GMF}) and in HG (regions limited by full lines, the field configuration with $y_0=0$ and $\theta_0=80^\circ$).
For parameters in the shaded regions (\red blue \black for $E=18$~TeV, pink for $E=251$~TeV, \red violet \black for both energies), the mixing is strong in both galaxies. Other notations are the same as in Fig.~\ref{f:probplot}.
}
\end{figure}
\subsection{From strong mixing in HG to ALP interpretations of GRB~221009A observations}
\label{sec:concl:interp}
Here, we estimate the ALP-photon mixing in HG only. To assess whether high-energy photons from GRB~221009A may indeed be related to ALPs, one needs (i)~to calculate the probability of the mixing along the entire way from the source to the observer, and (ii)~to compare the predicted photon spectrum, taking the potential impact of ALPs into account, with observations. The point (i) was addressed in various works, e.g.\ \cite{Roncadelli-newGRB,ST-GRB-JETPL,Meyer-newGRB,Roncadelli-assessment} etc. Mixing in the intergalactic space is expected to be small, and the resulting probability is determined by the conversion in HG, calculated here, and reconversion in the Milky Way. \black Note that, because of the low Galactic latitude of GRB~221009A, $b\approx 4^\circ$, the line of sight passes through the disk of our Galaxy as well, providing conditions for strong conversion. Figure~\ref{f:HG-MW} demonstrates that the strong conversion in both galaxies is expected for large regions of the parameter space.

\black However, the point (ii) requires more detailed information about the GRB flux and spectrum at highest energies. Presently, this information is unavailable since the GRB~221009A flux from Carpet-2 has not been published yet\redn, nor a widely accepted model of the GRB emission above 10~TeV exists, which would be necessary to estimate the intrinsic spectrum, and hence the amount of absorption. \black Therefore, we focus here only on one \black important \black element of the scenario and postpone a detailed quantitative interpretation of the observations to future work. \black It remains to be understood whether the ALP mechanism is necessary to explain the observed gamma-ray spectra\redn, and whether the strong mixing at both LHAASO and Carpet-2 energies is required. \black If interpretation of the spectra will not require nonstandard physics, one could use the model developed here to constrain ALP parameters from these observations. \black 

To put ALP parameters, for which we expect strong mixing in HG, in context, note that there are three classes of general constraints on $m$ and $g$ \cite{PDG2022}. Firm model-independent bounds from purely laboratory experiments are weak, and the entire parameter space in Fig.~\ref{f:probplot} is allowed by them. \red Note, however, that the relevant range of $g$ and $m$ is expected to be tested by the ALPS-II experiment \cite{ALPS-II} in future. \black The second group consists of constraints related to production of ALPs in stellar interiors; it is more model-dependent but still considered robust. It includes direct search of solar axions with laboratory experiments on the Earth and constraints from stellar evolution. For light ALPs we discuss here, both the solar \cite{CAST} and stellar \cite{HBstars} strongest upper limits on $g$ coincide occasionally and do not depend on $m$. A large part of the $(m,g)$ space, for which the strong mixing in HG is expected, corresponds to values of $g$ satisfying this limit. Finally, a number of astrophysical constraints are based on the effects of ALP-photon mixing, and therefore have to assume certain magnetic fields in the sources. These constraints are more indicative than robust because cosmic magnetic fields are rarely known precisely \black\cite{PDG2022}\black. The strongest \red claimed \black bound of this kind comes from observations of polarized magnetic white dwarfs \cite{MWD-polarization}. \redn For somewhat larger ALP masses, Ref.~\cite{polar-cap} claims a strong bound assuming the polar-cap model of pulsar emission. They are
\black
in conflict with various positive claims suggesting that ALPs with parameters disfavored by this constraint may help to solve certain discrepancies in stellar energy losses and in some gamma-ray observations of various extragalactic and Galactic objects, \black see e.g.\ reviews \cite{ST-mini-rev,Roncadelli-review2022,PDG2022}. \
\black 

Taken at the face value in Fig.~\ref{f:probplot}, this bound disfavours the parameters where the maximal mixing is most probable in HG, leaving favoured the larger-$m$ part. However, 
\red
systematic uncertainties related to the use of particular field models and to estimates of astrophysical contributions to polarization do not allow us to treat this bound as an exclusion in the particle-physics sense. Moreover, the strong mixing in HG is in fact possible for $(m,g)$ satisfying even this bound:
\black
recall the probabilistic interpretation of Fig.~\ref{f:probplot}, see Sec.~\ref{sec:ALP-gamma:unknowns}. The strong mixing in HG is possible for a wide range of ALP parameters (dark shade), for both LHAASO-KM2A and Carpet-2 energies, in more than 68\% of realizations of unknown parameters $\theta_0$ and $y_0$. Outside of the contours, the strong mixing is still possible for many\red, though less than 68\% of, \black realizations, and it is unknown, which particular $(\theta_0,y_0)$ are realized in HG. 
\black
This is illustrated in Fig.~\ref{f:HG-MW}, where strong-mixing contours for a particular realisation of parameters discussed in Sec.~\ref{sec:ALP-gamma:unknowns} extend to higher $m$ and lower $g$ compared with 68\% CL contours in Fig.~\ref{f:probplot}.
\black

\subsection{Summary}
\label{sec:concl:summary}
Making use of the first \black Hubble Space Telescope \black observations of the host galaxy of GRB~221009A, we construct a toy model of its magnetic field and calculate the probability of the photon-ALP conversion there for gamma-ray energies \black in the LHAASO and Carpet-2 bands. \black Thanks to the specific edge-on orientation of the host galaxy, the conversion is strong for a large part of the ALP parameter space because of the long path of a photon in the disk magnetic field, for both \black bands\black. In the Milky Way, the GRB was also seen through the disk, which opens up the possibility of efficient photon-ALP conversion in the host galaxy and reconversion in the Milky Way. This could \black help to \black explain the detection of anomalously energetic photon-like events from the GRB\black, if more detailed analysis of the yet unpublished spectra would support the anomaly. If not, the conversion model presented here might be used to constrain ALP parameters\black.

\appendix
\section{Expressions for the field model}
\label{a:field}
This supplementary section collects explicit expressions for calculation of the magnetic field in the host galaxy of GRB~221009A within the toy model constructed in Sec.~\ref{sec:field:model}. See the main text and \cite{Pshirkov-GMF,Xshape} for explanations. Numerical values of parameters are collected in Table~\ref{tab:param}.

We introduce the cylindrical coordinate system $(r,\theta,z)$ with the origin in the host galaxy center. Here, $r$ is the radial distance in the disk plane, $\theta$ is the angle in the disk plane, and $z$ is the height above the disk plane. The angle $\theta_0$ measures the orientation of the overall spiral structure with respect to the line $\theta=0$; conversion probabilities in the main text were averaged over $\theta_0$. The disk and toroidal halo field components are set to 0 at $r^2+z^2 >R_{\rm lim}^2$. Dimensionful field components are denoted as $(B_r,B_\theta,B_z)$.
\subsection{Disk field}
\label{a:field:disk}
\red These expressions describe the disk field in the parametrization of Ref.~\cite{Pshirkov-GMF}: \black
\[
b(r)=
\left\{
\begin{array}{lc}
B_0 \frac{R}{R_{\rm c}\cos\theta_0,}    & r<R_{\rm c}, \\
B_0 \frac{R}{r\cos\theta_0,}    & r \ge R_{\rm c};
\end{array}
\right.
\]
\[
B_{\rm d}(r,\theta,z)= 
\left\{
\begin{array}{lc}
\! b(r) \cos \left( \theta- \frac{1}{\tan p} \ln \frac{R_{\rm c1}}{R} + \theta_0 \right) \mbox{e}^{-|z|/z_0} , \!  & r<R_{\rm c1}, \\
\! b(r) \cos \left( \theta- \frac{1}{\tan p} \ln \frac{r}{R} + \theta_0 \right) \mbox{e}^{-|z|/z_0} , \!  & r\ge R_{\rm c1}.
\end{array}
\right.
\]
\subsection{Toroidal halo field}
\label{a:field:toro}
\red These expressions describe the halo field in the parametrization of Ref.~\cite{Pshirkov-GMF}: \black
\[
B_{\rm H}(r,\theta,z)=
\mbox{sign}(z) B_{\rm 0H} 
\left[ 
1+
\left(
\frac{|z|-z_{\rm 0H}}{z_{\rm 1H}}
\right)^2
\right]^{-1}
\!\!\!\!
\frac{r}{R_{\rm 0H}}
\exp \left( 1- \frac{r}{R_{\rm 0H}}  \right)\!,
\]
where
\[
z_{\rm 1H}=
\left\{
\begin{array}{lc}
z_{\rm 11H},    & r<R_{\rm c}, \\
z_{\rm 12H},    & r\ge R_{\rm c}.
\end{array}
\right.
\]
\subsection{X-shaped halo field}
\label{a:field:Xshape}
\red These expressions describe the additional component of the halo field from Ref.~\cite{Xshape}: \black
\[
r_1(r,z)=1/(1+a z^2);
\]
\[
B_{\rm c}(r,z)=B_1 \exp \left( -r_1(r,z) \frac{r}{L_{\rm X}}\right);
\]
\[
B_{r \rm X}(r,z)=2ar_1(r,z)^3 r z B_{\rm c}(r,z);
\]
\[
B_{z \rm X}(r,z)=
\left\{
\begin{array}{lc}
r_1(R_{\rm c1},z)^2 B_{\rm c}(R_{\rm c1},z),    & r<R_{\rm c1}, \\
r_1(r,z)^2 B_{\rm c}(r,z),     & r\ge R_{\rm c1}.
\end{array}
\right.
\]
\subsection{Total field components}
\label{a:field:comp}
\red These are the components of the total field in our model, in cylindrical coordinates: \black
\[
B_r (r,\theta, z)=B(r,\theta,z) \sin p + B_{r \rm X}(r,z);
\]
\[
B_\theta (r,\theta, z)=B(r,\theta,z) \cos p + B_{\rm H}(r,\theta,z);
\]
\[
B_{z}(r,z)=B_{z \rm X}(r,z).
\]

\begin{table}
\centering
\begin{tabular}{ccccc}
\hline\hline
Parameter & Value &~~~~~& Parameter & Value\\
\hline
$B_0$ & 4.39~$\mu$G && $z_0$&0.49~kpc \\
$B_{\rm 0H}$& 2.19~$\mu$G && $R_{\rm 0H}$& 3.92~kpc\\ 
$B_1$& 2.19~$\mu$G && $z_{\rm 0H}$&0.637~kpc\\
$p$&$-6^\circ$ && $z_{\rm 11H}$&0.1225~kpc\\ 
$d$&$-$0.294~kpc && $z_{\rm 12H}$&0.196~kpc\\
$R$&4.165~kpc && $R_{\rm c1}$&1.225~kpc\\
$R_{\rm c}$&2.94~kpc && $a$&0.065~kpc$^{-2}$\\
$R_{\rm lim}$&9.8~kpc && $L_{\rm X}$&14.7~kpc\\
\hline
\end{tabular}
\caption{\label{tab:param}
Parameters of the field model.}    
\end{table}
\acknowledgments
The author thanks Dmitry Gorbunov, Maxim Pshirkov, Grigory Rubtsov and Peter Tinyakov for discussions.
This work is supported by the RSF grant 22-12-00253.

\red
\paragraph{Note added.} 
After the submission of the present paper, the LHAASO collaboration published a preprint \cite{LHAASO-KM2A-GRB} with some additional KM2A data, which were subsequently used by Gao et al.~\cite{LHAASO-ALP-limits} to constrain some ALP parameters with the help of the HG magnetic-field model constructed here.
\black
\bibliographystyle{JHEP}
\bibliography{grb-host}

\label{lastpage}
\end{document}